# mrPUF: A Memristive Device based Physical Unclonable Function

Omid Kavehei[†], Chun Hosung[†], Damith Ranasinghe[‡], Stan Skafidas[†]
[†]Department of Electrical and Electronic Engineering, The University of Melbourne, Victoria 3010, Australia
[‡]Auto-ID Labs, The University of Adelaide, SA 5005, Australia
Emails: {omid.kavehei,hosung.chun,sskaf}@unimelb.edu.au, damith@cs.adelaide.edu.au

***Abstract*—Physical unclonable functions (PUFs) exploit the intrinsic complexity and irreproducibility of physical systems to generate secret information. PUFs have the potential to provide fundamentally higher security than traditional cryptographic methods by preventing the cloning of identities and the extraction of secret keys. One unique and exciting opportunity is that of using the super-high information content (SHIC) capability of nanocrossbar architecture as well as the high resistance programming variation of resistive memories to develop a highly secure on-chip PUFs for extremely resource constrained devices characterized by limited power and area budgets such as passive Radio Frequecy Identification (RFID) devices. We show how to implement PUF based on nano-scale memristive (resistive memory) devices (mrPUF). Our proposed architecture significantly increased the number of possible challenge-response pairs (CRPs), while also consuming relatively lesser power ($\approx$ 70 $\mu$W). The presented approach can be used in other silicon-based PUFs as well.**

## I. Introduction

Physically unclonable functions (PUFs) are innovative physical security primitives that expresses an inherent and unclonable instance-specific feature of a physical object. The most noteworthy property of a PUF is its *unclonability*. A PUF produces an output signal (*response*), which is a function of physical properties of the system, and possibly of an external physical excitation signal (*challenge*). It has to satisfy several characteristics including: (i) a sufficiently large margin between expected responses to the same challenge on two distinct instances of PUFs to avoid collision of responses. (ii) sufficiently small difference between two separately measured responses to the same challenge on a single random instance of a PUF, and (iii) sufficiently large number of independent (information) bits that can be generated by a PUF.

There are different types of PUFs. A comprehensive review of different kind of PUFs can be found in [1]. This paper introduces a memristive device-based PUF (mrPUF) using delay-based ring-oscillator (RO-) and super-high information content (SHIC-) PUF concepts. The main idea of a SHIC-PUF is proposed in [2] based on the unprecedented amount of device density ($10^{10}$ bit/cm$^2$) in crossbar arrays. This idea is based on the Readout time variation of a memory cell consisting of a single diode, which makes this approach extremely slow (100 bit/s). This drawback dramatically reduces the practicability of the SHIC-PUF in real world applications. Another technique for achieving SHIC is monolayer based Public PUF (PPUF), which is a PPUF based on a network of randomly distributed metallic particles connected by self-assembled molecules (devices) [3].

In our proposed structure, nanocrossbar memories consist of resistive memory (memristive) devices can be used together with established PUF structures to enhance reliability, unpredictability, energy saving and area requirements. This paper proposes a current-controlled ring-oscillator PUF, where its delay is controlled not only by CMOS and nanocrossbar (array of nanodevices) manufacturing variations but also random selection of $C$ rows out of $N$ rows within a $N \times M$ crossbar. CMOS compatibility, highly dense, potentially low-cost, relatively high defect (stuck-at-ON state or stuck-at-OFF state) rate, significantly high variation, sneak current path issue, and perfect tolerance against electromagnetic radiation and high-energy cosmic particles (compare to CMOS flip-flops) of memristive devices in general and resistive random access memories (RRAMs) in particular, make them a promising candidate for realizing PUFs [4]–[6]. Our proposed architecture satisfies the third property of PUFs in a way that allows significant margins (in terms of invalid pairs) for the first and second properties, because of unprecedented amount of possible configurations. In this paper, we compare our mrPUF with the state-of-the-art RO-PUF design in terms of total number of challenge-response pairs (CRPs), without considering the impact of invalid pairs.

The paper is organized as follows: Section II discusses original and state-of-the-art RO-PUFs. Proposed architecture and its CRP analysis is presented in Section III, and in Section IV we conclude the paper.

## II. Conventional, Configurable and multi-$V_{dd}$ RO-PUF

### A. Conventional RO-PUF

Identical ring-oscillators operate at different frequencies based on the magnitude of influence by device process variation. A RO-PUF consists of $k$ oscillators, two $k$-to-1 multiplexer that chooses a pair of ring-oscillators, $RO_i$ and $\text{RO}_j$, two counters and a comparator, as shown in Fig. 1. If $f_i < f_j$ therefore the output bit is 0, otherwise 1. Multiplexers' input is the challenge code.

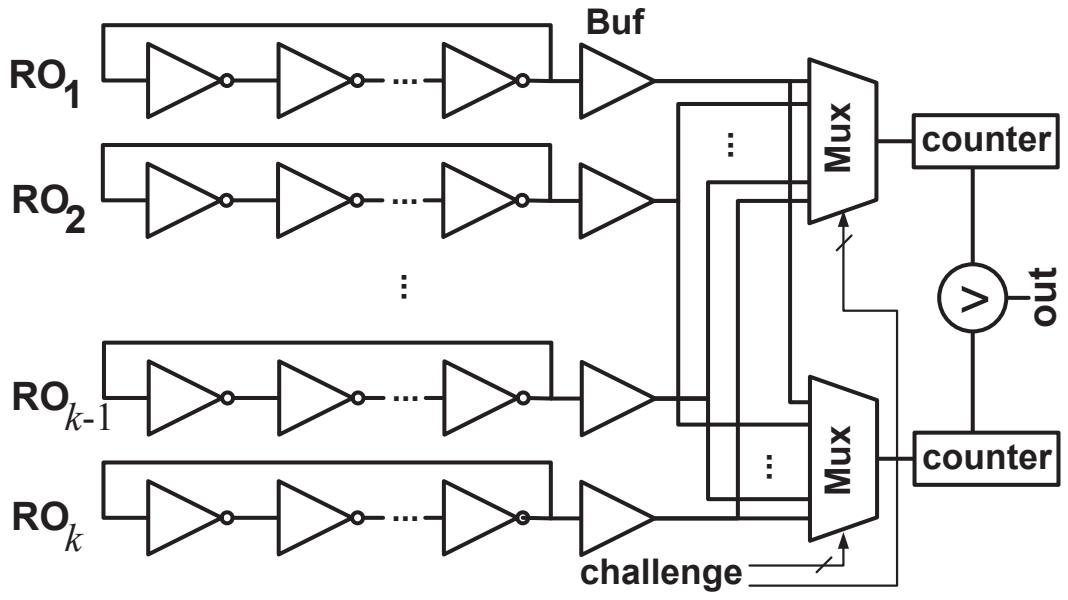

Fig. 1: Traditional RO-PUF. NAND gates that activate the ROs are not shown in this figure.

There are

$$\frac{k(k-1)}{2} \ , \qquad (1)$$

distinct pairs of ROs. However, entropy is not as much. There are some dependencies between ROs because their frequencies depend on process variation, if $f_i < f_j$ and $f_j < f_z$ then it is clear that $f_i < f_z$. Therefore, the entropy, or the maximum number of independent bits which is correspond to the maximum number of possible configurations, is much less ($k!$). The entropy in this case will be $\ln(k!)$ [7].

### *B. Configurable and multi-$V_{\text{dd}}$ RO-PUF*

There has been consistent efforts to improve the RO-PUF's reliability, uniqueness and resilience to attacks. An interesting example is a configurable RO-PUF [8]. This kind of RO-PUF configures the placement of ROs on chip using a multiplexer to increase the original RO-PUF's reliability. Another alternative is to insert multiplexers within the inverter chains to be able to configure the PUF's ROs interaction and structure. This method has shown a significant increase in the uniqueness of responses from 35.91% for the original RO-PUF to 45.51% for a configurable RO-PUF.

Statistically, adding more flexibility to the structure relaxes the RO-PUF's performance and responses are expected to be much less dependent on temperature and voltage variations because of increased number of configurations that an input challenge can make. Unsurprisingly, the number of unstable bits due to voltage variation is reduced to less than 5 out of a 127 bit response. For the original RO-PUF this number is $< 45$ bits. Recently, a multi-$V_{\text{dd}}$ ring-oscillator PUF (multi-$V_{\text{dd}}$ RO-PUF) was proposed which uses the Alpha law delay dependency of CMOS inverters on $V_{\text{dd}}$ [9]. The maximum number of CRPs is

$$\frac{k(k-1)}{2} \times L^C \ , \qquad (2)$$

where $L$ is the number of supply voltages (normally 2 or 3) and $C$ is the number of inverters in a chain. Although this method relaxes the selection flexibility more than the traditional RO-PUF, this approach is sensitive to voltage variation because it uses different voltage levels and requires appropriate minimum margins between different supply voltages. Also it might require level-shifters as $L$ increases. Therefore, not all limited number of CPRs are valid.

## III. Memristive Device based PUF

Recently, a number of unusual electronic phenomena have emerged that include resistive switching and memristive phenomena. Realization of a solid-state memristive device in 2008 [10], [11] is one of them and has created a new wave of research in realization of large memory arrays and data-intensive in-memory computing. A memristive nano-scale device is a non-volatile memory that is manufactured with a considerable amount of uncertainties. Besides manufacturing uncertainties, inter-die and intra-die variation, these devices have shown a significant programming variation in their resistance, which includes cycle-to-cycle variation and stochastic switching. Parameters like low state resistance ($R_{\text{ON}}$) and high state resistance ($R_{\text{OFF}}$), are random variables with log-normal distributions [6].

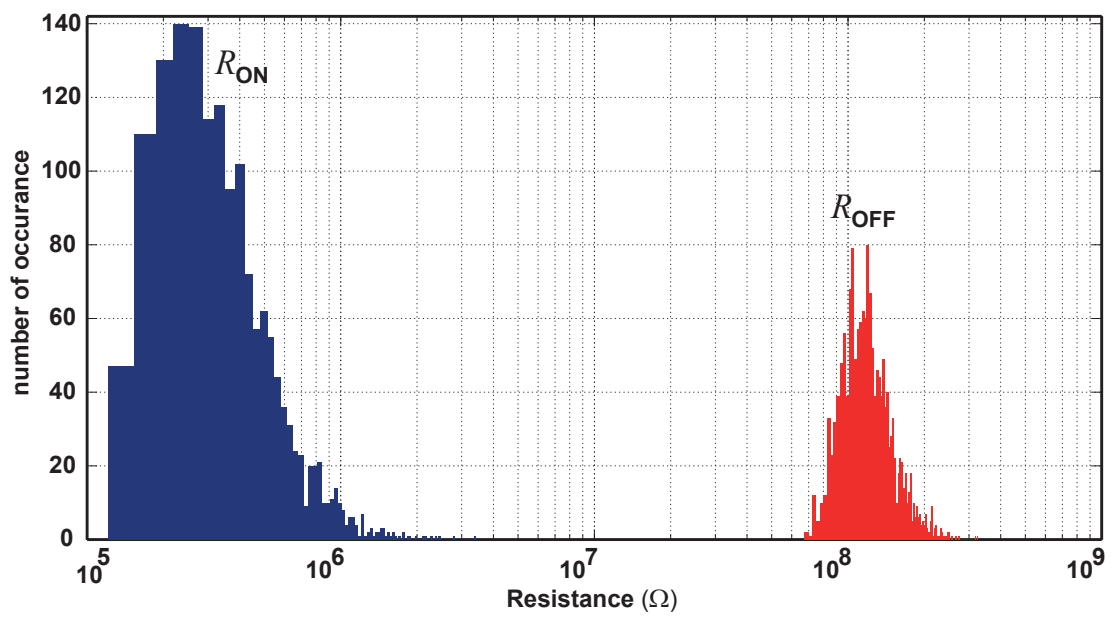

Fig. 2: Programming variation extracted from an experiment on a 40 × 40 nanocrossbar array (1600 crosspoint devices).

Fig. 3a shows the variation of high and low resistance after a random initial programming step with the same probability (0.5) for programming a cell as ON or OFF. In real-world applications, current compliance has to be set using a CMOS transistor. This model is in agreement with the experimental data in [6]. Due to a relatively high defect rate, we considered 10% of devices stuck-at-ON/OFF with 0.7% probability for stuck-at-ON and randomly distributed resistance values for both defects.

In a $N \times M$ nanocrossbar array, parallel devices in one row create total resistance of $R_{\text{total}}$. As demonstrated in Fig. 3b, this total resistor, through a series resistor, $R_{\text{ser}}$, is connected to a current mirror that controls delay of an inverter. This would create random configurations of currents depending on the total resistance value. Now, in a ring-oscillator, each stage (inverter) can be controlled with a total resistance of one row. Fig. 4 demonstrates a mrPUF system with $k$ ring-oscillators, $C$ stages of inverters and a $N \times M$ nanocrossbar array.

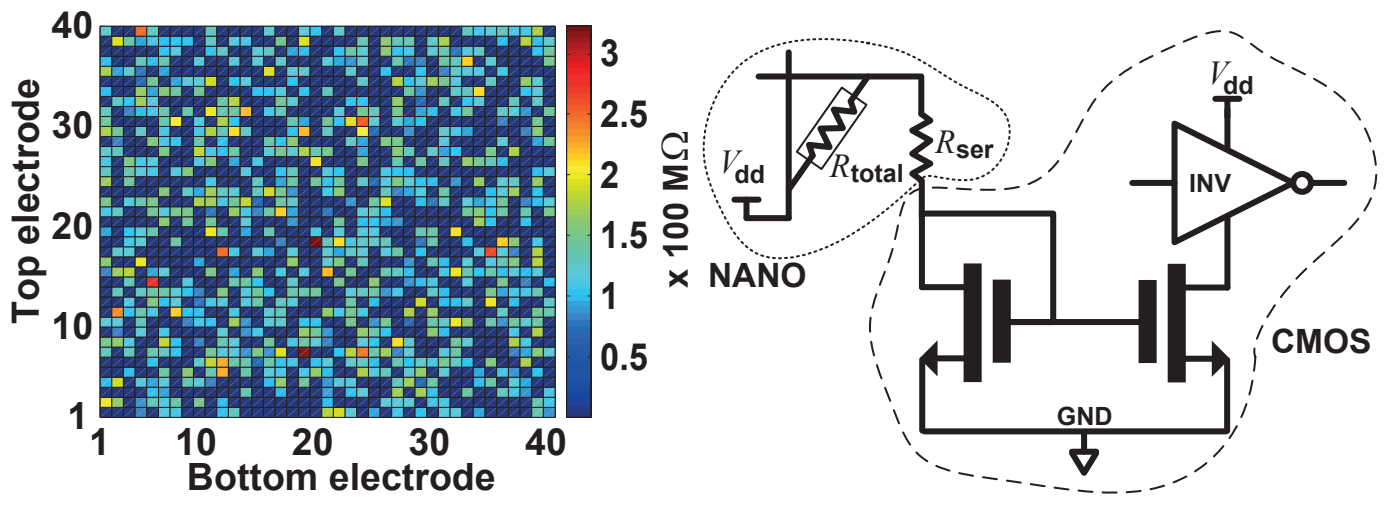

(a) (b)

Fig. 3: Initial programming variation of resistances across a nanocrossbar array can be used in mrPUF. (a) Programming variation across a 40 × 40 nanocrossbar. Each crosspoint is a resistive memory device that can be accessed and programmed using a top electrode and a bottom electrode. (b) A single current-controlled inverter cell of a mrPUF's ring-oscillator. Some of items in this cell are shared between ROs.

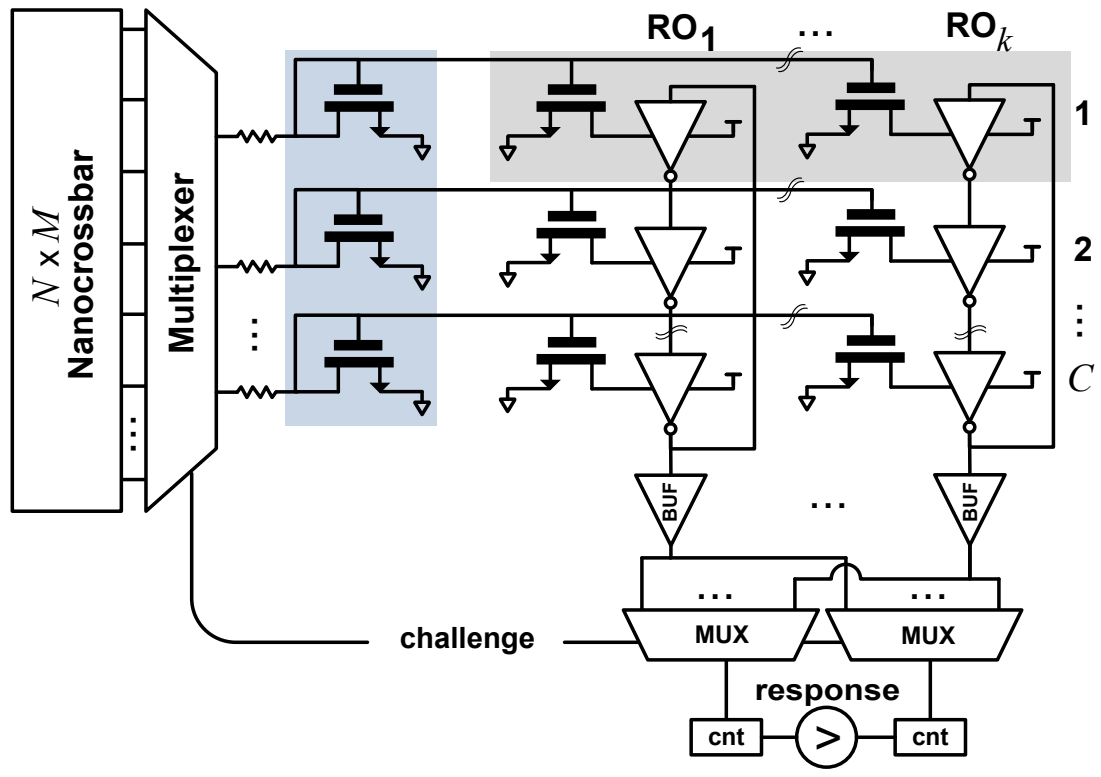

Fig. 4: mrPUF system. The highlighted areas shows global and local parts of the Readout CMOS circuitry.

In this situation, besides the CMOS process variation, number of ON/OFF devices and the randomness in their resistance values determine the ROs' frequencies. With the assumption of having at least one $R_{\mathrm{ON}}$ per row, the total number of different $R_{\mathrm{total}}$ values per row is $2^{M-1}$. Fig. 5 shows possible values for $R_{\mathrm{total}}$. Hence, with $C$-stage ROs, the number of CRPs for a mrPUF can be calculated as

$$\frac{k(k-1)}{2} \times \frac{N!}{(N-C)!} \times 2^{C(M-1)} \ , \qquad (3)$$

where $N!/(N-C)!$ is implemented by the multiplexer next to the nanocrossbar array in Fig. 4. The maximum number of CRPs for a mrPUF and a multi-$V_{\mathrm{dd}}$ RO-PUF with $k = 4$, $N = M = 10$, $L = 3$ and $C = (2, 4, ..., 10)$ (and a NAND gate), can be seen in Fig. 6. This is a significant jump in the number of possible CRPs. A mrPUF system with $N = M = 40$ consumes 65 $\mu$W to 74 $\mu$W (this data is extracted from several mrPUFs) and produce significantly higher number of CRPs.

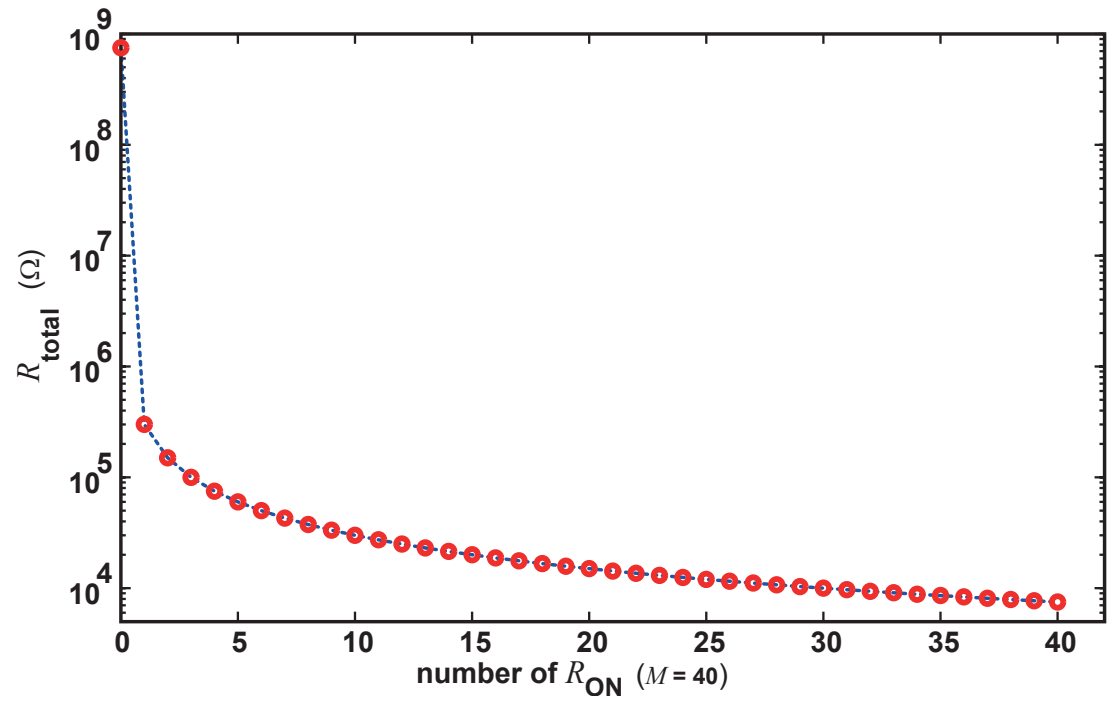

Fig. 5: $R_{\mathrm{total}}$ versus the number of $R_{\mathrm{ON}}$ resistances.

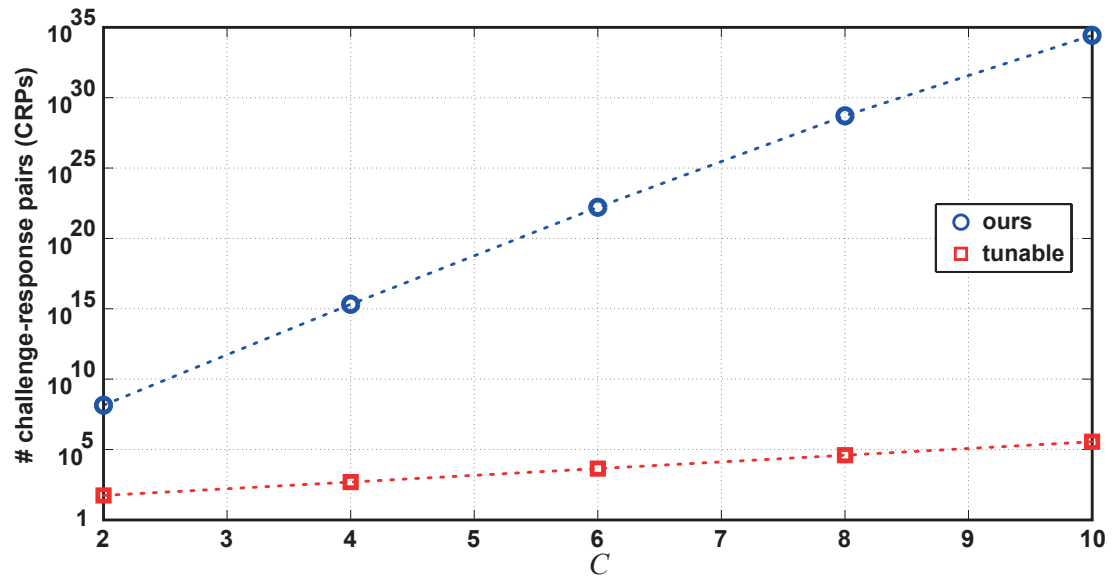

Fig. 6: Maximum CRPs of the mrPUF against the multi-$V_{\mathrm{dd}}$ RO-PUF.

Considering requirements of a conventional RO-PUF to generate reasonably large number of CRPs, this is a significant increase utilizing only three CMOS inverters. All possible CRPs are not valid. A careful margin analysis for voltage and temperature variation should be taken into account and some techniques, like configurable RO-PUF could improve performance of mrPUFs. Our simulations is conducted at 70° Celsius for the CMOS devices and no temperature consideration for the nanodevices. Besides, long term stability of the resistive states is required, thus devices offering good retention and less degradation of resistive values are preferable. In addition, further analysis is required using experimental results from [12] and [13] to show the number of valid CRPs under more realistic conditions, such as temperature variation. There works show that electrical conductivity of these devices and amplitude of variation –due to Joule heating– follow $1/e^{-T}$ relationship with temperature.

Fig. 7 illustrates the sensitivity of the output frequency to the $R_{\mathrm{total}}$ variation. RO1 shows the result when $R_{\mathrm{total}}$ values are fixed. RO2 represents a situation where $R_{\mathrm{total}}$ variation is assumed but two $R_{\mathrm{total}}$ values have the same mean values. If we randomly assign values for $R_{\mathrm{total}}$ and consider the same variation we observe that the output frequency is not only a function of CMOS process variation but also a function of $R_{\text{total}}$ configuration and variation (See RO3 in Fig. 7).

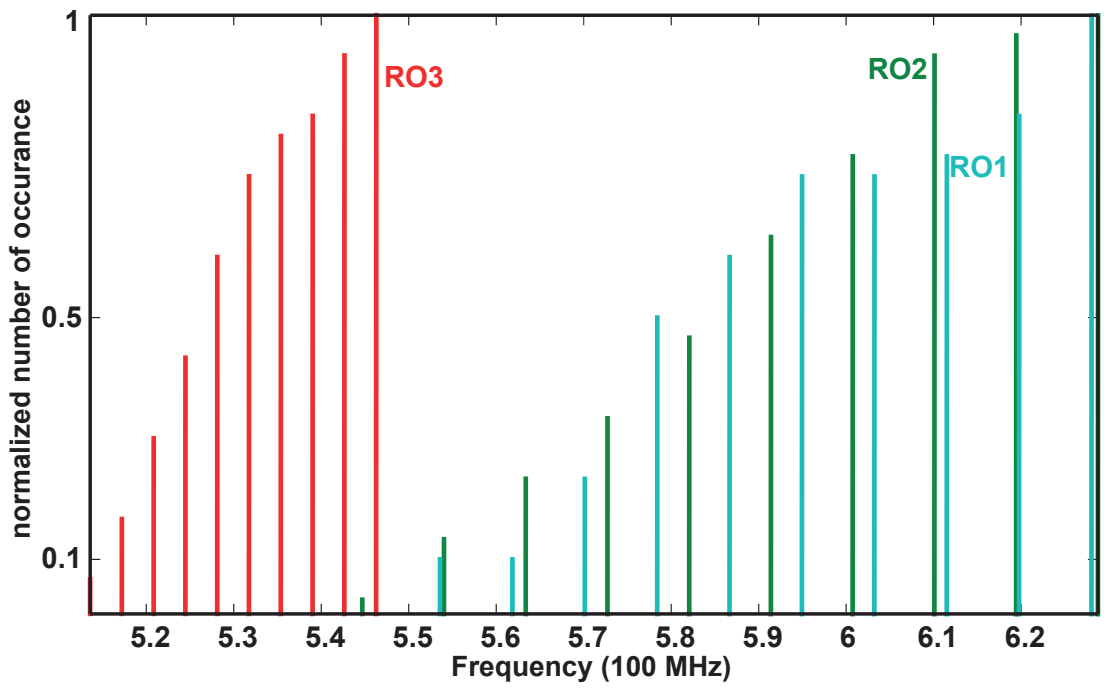

Fig. 7: A cumulative histogram of output frequencies for 3 ring-oscillators consist of two-stage inverters (that are current-controlled by two $R_{\text{total}}$) and a NAND gate.

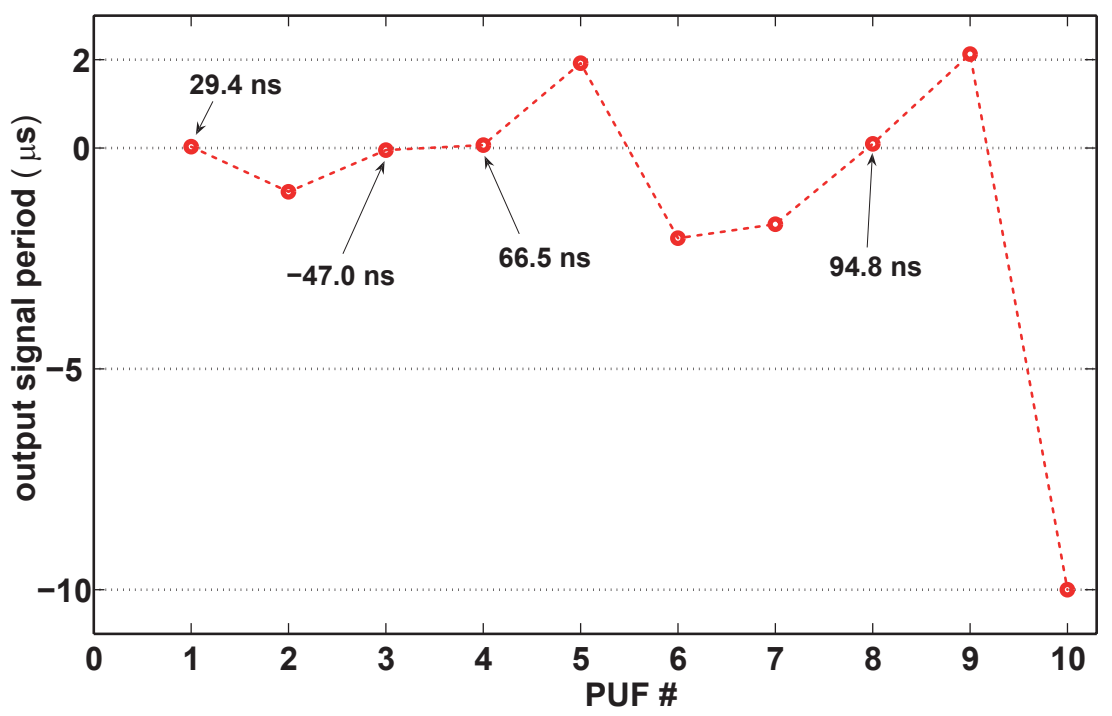

Fig. 8: Output signal period of 10 mrPUF. This figure shows one bit output per mrPUF.

We conducted our simulations using two ($k = 2$) ROs with 10 stages ($C = 10$) and a $40 \times 40$ nanocrossbar array with 1.25 $\Omega$ segment resistance for nano-wires. The maximum number of challenge-response pair is $7.75 \times 10^{132}$. Series resistor, $R_{\text{ser}}$, is 1 k$\Omega$. Each nanodevice is a resistive memory that is programmed randomly as $R_{\text{ON}}$ or $R_{\text{OFF}}$ with the given variation in Fig. 2 and the mentioned defect distribution and ON and OFF programming probability. We considered only one-time programming, so the system is not assumed to be re-programmable. Readout is done using a lower voltage (0.6 V) than programming threshold of the device. In simulations, we used IBM 65 nm CMOS technology with a 0.6 V as supply voltage. We simulated 10 different instances of mrPUF and apply similar challenge to all of them and then extracted response bits which is the result of a comparison between frequencies of the two ROs. Results for one bit is shown in Fig. 8. A uniqueness ($U$) analysis using SPICE simulation on a 10-bit output shows $U = 50\%$ which is an ideal outcome.

## IV. Conclusion

Unlike other silicon-PUFs that are based on manufacturing variability of CMOS devices, we considered not only the uncertainties during manufacturing but also randomness in super-high information content (SHIC) that the emerging nanocrossbar architectures are offering. This approach enabled us to propose an authentication method, which promises unprecedented number of possible challenge-response pairs compare to conventional PUFs. This paper introduced a readout technique that utilities overall resistance variation of non-volatile resistive memory (RRAM) devices to tune delay of a current-controlled ring-oscillator. Controlling this randomness for conventional computing is a difficult and costly task. This extremely large process and programming variation of nanodevices comes for free and is plentiful The mrPUF turns this disadvantage into an advantage.

## Acknowledgements

The authors would like to thank Dr Eike Linn, from RWTH-Aachen, for his insightful comments on resistive memories and overall quality of the paper.